# Полет над квантовой хромодинамикой

З.К. Силагадзе

Институт Ядерной Физики им. Г.И. Будкера и Новосибирский Государственный Университет

## Часть вторая: Квантовая механика (Первую часть можно посмотреть здесь).

Нормальные люди (т.е. не физики) обычно квантовую механику представляют как предмет чрезвычайно туманный и далекий от здравого смысла. И частично это действительно так. Вот знаменитая история об известных математиках Расселе и Уайтхеде (из книги М. Кац, Несколько вероятностных задач физики и математики. Москва, Наука, 1967):

"Один из них делал доклад, а другой был председателем. Скажем, Рассел был председателем, а Уайтхед - докладчиком. Доклад был посвящен основам квантовой механики. Не только слушатели были уже по горло сыты этим докладом, но и председатель. Все было исключительно трудно, путанно и неясно. Когда доклад закончился, председатель почувствовал, что он должен как-то этот доклад прокомментировать. И вот он сказал только одну фразу, которая одновременно была и вежливой, и правдивой. Он просто сказал: «Мы должны быть благодарны докладчику за то, что он не затемняет далее этого и так уже достаточно запутанного предмета»".

Тем не менее, наше ощущение логической простоты классической механики и запутанности квантовой, обманчиво. Оно, в значительной степени, результат нашего воспитания и многолетней учебы в духе западной культурной традиции. Но если захотим спокойно разобраться, что же стоит за основными понятиями классической механики, такими как масса, сила, инерция, не говоря о пространстве и времени, нас ждет разочарование (см., например, Э. Мах, Механика. Историко-критический очерк ее развития. Москва, Эдиториал УРСС, 2000) – все не менее запутанно и неясно. Наше традиционное обучение вначале направлено на то, чтобы привить человеку объективность классической физики. Но "объективность классической физики - что-то вроде полуправды [1]. Она весьма хороша, представляет собой выдающееся достижение, но почему-то затрудняет полное понимание реальности в гораздо большей степени, чем кажется." Поэтому потом приходится эти классические понятия ломать. Весьма странный способ обучения, не так ли?

На самом деле, логические основы квантовой механики просты и прозрачны. Представим себе, что электрону надо перейти из одной точки пространства-времени в другую. Как же он это сделает? Важно осознать [2], что у электрона нет мозгов, он слишком мал для этого. Следовательно, ему очень трудно выбрать свой путь из огромного множества всевозможных путей. Единственная подходящая директива

для него будет:

- Используй ВСЕ пути!

При этом все пути, которые можно только вообразить, соединяющие две данные точки, равноправны. Но если нет никакой разницы между путями, электрон вряд ли куда-нибудь придет. Как же сделать, чтобы пути были равноправны, но тем не менее отличались? У электрона нет мозгов, но у нас то есть! Сопоставим каждому пути комплексное число (см. Примечание А) одного и того же модуля (пусть единичного), а фаза этого числа пусть зависит от пути. Просуммируем эти числа по всем путям, чтобы получить так называемую амплитуду перехода. И так, второе основное правило:

- Каждый путь $P$ входит с множителем $z(P) = e^{is(P)}$. Амплитуда перехода равна сумме $z(P)$ по всем путям.

Полученная таким образом, амплитуда перехода из точки $x_1$ в точку $x_2$, $K(x_1, x_2)$, будет комплексным числом, но его модуль уже не будет равняться единице. Величина этого модуля и определяет вероятность обнаружить электрон в точке $x_2$ согласно третьему основному правилу:

- Вероятность обнаружить электрон в точке $x_2$ равна $\mid K(x_1, x_2) \mid^2$.

Вот и все. Хотите верьте, хотите нет (а еще лучше посмотрите первоисточник (например книгу Р. Фейнман, Г. Хиббс, Квантовая механика и интеграл по траекториям. Москва, Мир, 1968), эти три простые правила как раз и составляют основу квантовой механики.

Но как бы я не старался усыпить ваш, испорченный западным рационализмом, здравый смысл, вы наверное все равно почувствовали неладное. Они, эти правила, возможно, и действительно простые, но если следовать им неукоснительно, странных следствий не избежать.

Во-первых, от них так и веет идеализмом, и выше я это намеренно подчеркнул. Второе и третье правило вряд ли под силу безмозглому электрону, требуется участие нашего сознания. Классическая физика привыкла не принимать сознание во внимание, называется это материализмом. Но, если задуматься, еще неизвестно что более естественно: считать что электрон "умеет" решать дифференциальные уравнения, чтобы следовать законам Ньютона, или принять более активную роль сознания в формировании реальности.

Что такое объективная реальность? На этот вопрос не так легко ответить, как может показаться. Иначе не было бы философии. Иногда сознание рассматривают просто как зеркало реальности. Как бы не так!

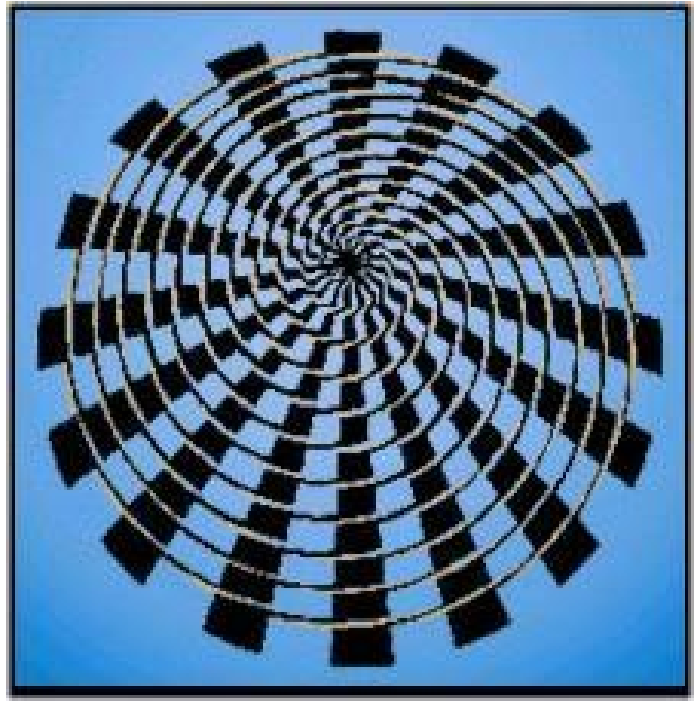

Посмотрите на рисунок. Видите ли спираль на этом рисунке? Отчетливо ли видите? Так вот, никакой спирали там нет, одни окружности! Если не верите, попробуйте проследить эти окружности, начиная с нижней точки, или посмотрите аналогичный рисунок здесь и мышкой поместите стрелку-указатель в пределах рисунка. Теперь обратимся к другому рисунку (Надо перейти по ссылке, так как в pdf файле нет анимации. Другой вариант этого замечательного эффекта можно посмотреть здесь [3]).

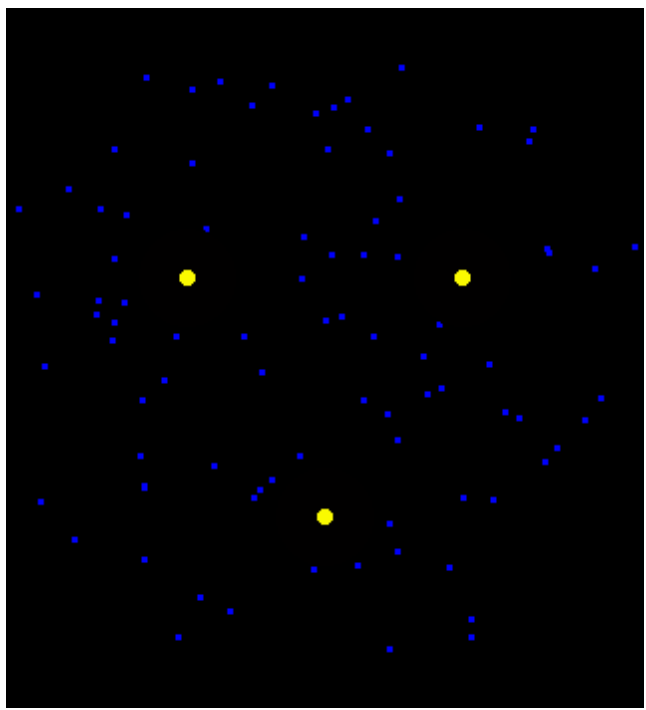

Зафиксируйте внимание на подвижные голубые точки. Через некоторое время неподвижные желтые точки станут исчезать! Все это трюки нашего сознания. Он формирует ощущаемый нами мир весьма нетривиальным образом, и мы можем даже не подозревать на что еще он способен. Как сказал английский поэт и мистик Ульям Блэйк, "Дурак видит не то же самое дерево, что видит умный".

Я вовсе не утверждаю, что квантовая механика обязательно требует участия сознания, хотя подобную мысль высказывал не кто иной, как сам великий Вигнер (см. Е. Вигнер, Вероятность существования самовоспроизводящейся системы. В книге: Этюды о симметрии, Москва, Мир, 1971). Известно аж девять интерпретаций квантовой механики [4], а недавно появилась еще одна [5] (см. еще вот эту [6], и эту[7]). Из них только две апеллируют к сознанию, но если вы думайте, что они самые экзотичные, то ошибаетесь [8].

Но вернемся к странным следствиям квантовой механики. Вероятность обнаружить электрон отлична от нуля для многих точек, т.е. электрон как бы одновременно присутствует во многих точках. Не странно ли это? Более в общем случае, это свойство квантовой механики приводит к принципу суперпозиции: как правило, если квантовая система может находится в двух разных состояниях, то линейная суперпозиция этих состояний тоже является допустимым состоянием системы. Но как можно представить "линейную суперпозицию" двух состояний? Для привычных нам макроскопических тел это довольно трудно, как показывает следующий пример Шредингера [9](см. также здесь [10]):

"В стальной камере заключена кошка, вместе со следующим дьявольским устройством (которого следует от кошки надежно экранировать): в гейгеровском счетчике помещена крупинка радиоактивного вещества, такой величины, что в течение часа один из атомов может распасться, но с равной вероятностью может и нет. Если распад произойдет, гейгеровский счетчик срабатывает и через реле освобождает молоток, который разбивает капсулу с синильной кислотой. Если эту систему предоставить самому себе в течение часа, то можно сказать, что кошка все еще жива, если ни один атом не распался за это время. Волновая функция всей системы выразила бы это, имея в себе живую и мертвую кошку (извините за выражения) смещенных или размазанных в равной степени." Полная бессмыслица с точки зрения кошки.

Но то, что является странным и абсурдным в макроскопическом мире, в микромире может оказаться самим обычным делом. Более того, наше с вами существование, скорее всего, основано как раз на квантовомеханическом смешивании частиц (см. Примечание В). По современным представлениям, масса частиц возникает за счет взаимодействия частицы с вакуумом. Называется это механизмом Хиггса.

Британский физик Дэвид Миллер придумал очень остроумное объяснение [11] что такое механизмом Хиггса. Приведем его в несколько видоизмененом в виде. Представим себе банкет физиков после открытия бозона Хиггса. Банкетный зал и физики – это модель вакуума, который не является пустотой, а заполнена скалярным полем (физиками). Если хотите, можете назвать это заполняющее все пространство скалярное поле эфиром. Но название «скалярное» означает, что оно одинаково во всех инерциальных системах отсчета и существование такого «эфира» не противоречит теории относительности.

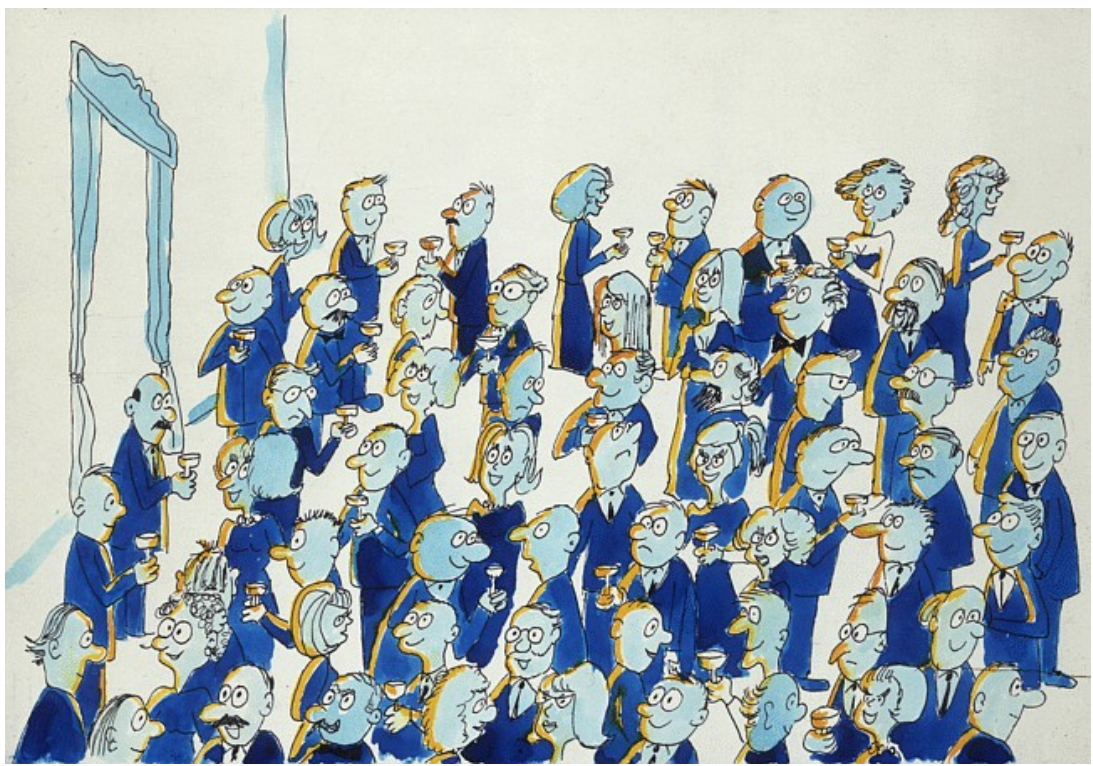

Вдруг в зал входит очень красивая женщина и вокруг нее моментально собирается группа поклонников.

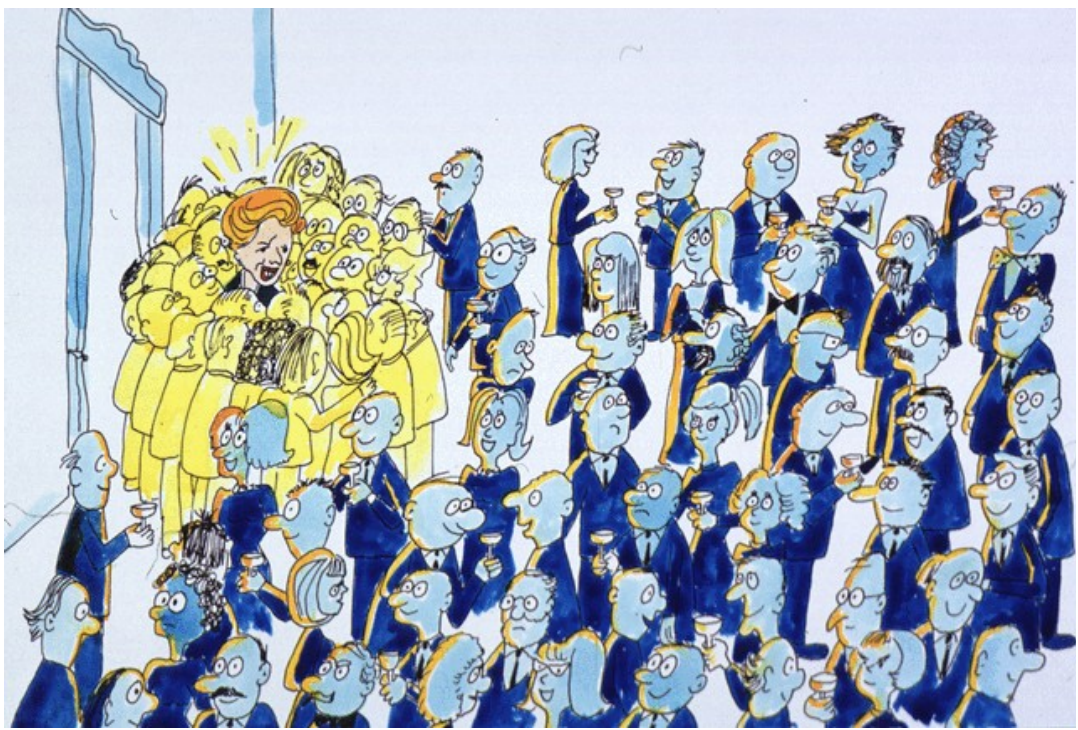

Этот кластер передвигается по залу и его гораздо труднее остановить, чем женщину. Это механизм Хиггса – эфемерное создание за счет поклонников (взаимодействие с вакуумом) приобретает значительную массу.

Женщина выходит из зала. Но кто-то распространяет сплетню про нее.

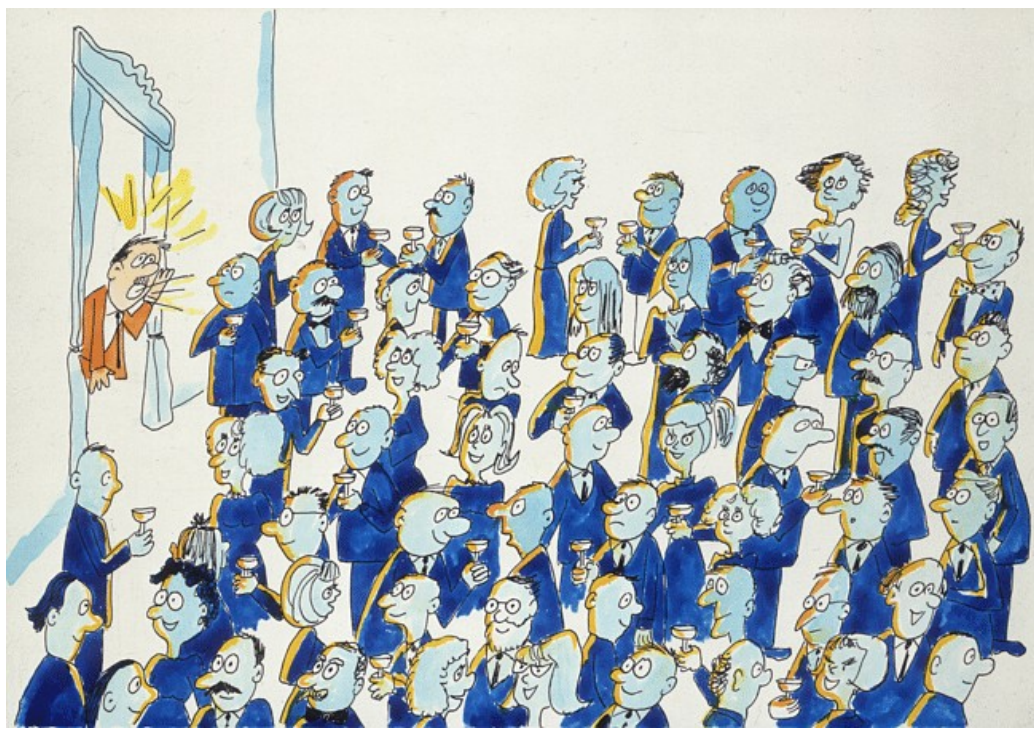

Падкие на сплетню физики сбиваются в кучу и обсуждают сплетню. Эта группа сплетников, которая тоже имеет большую массу и перемещается по залу, является аналогом бозона Хиггса (возбуждение скалярного поля).

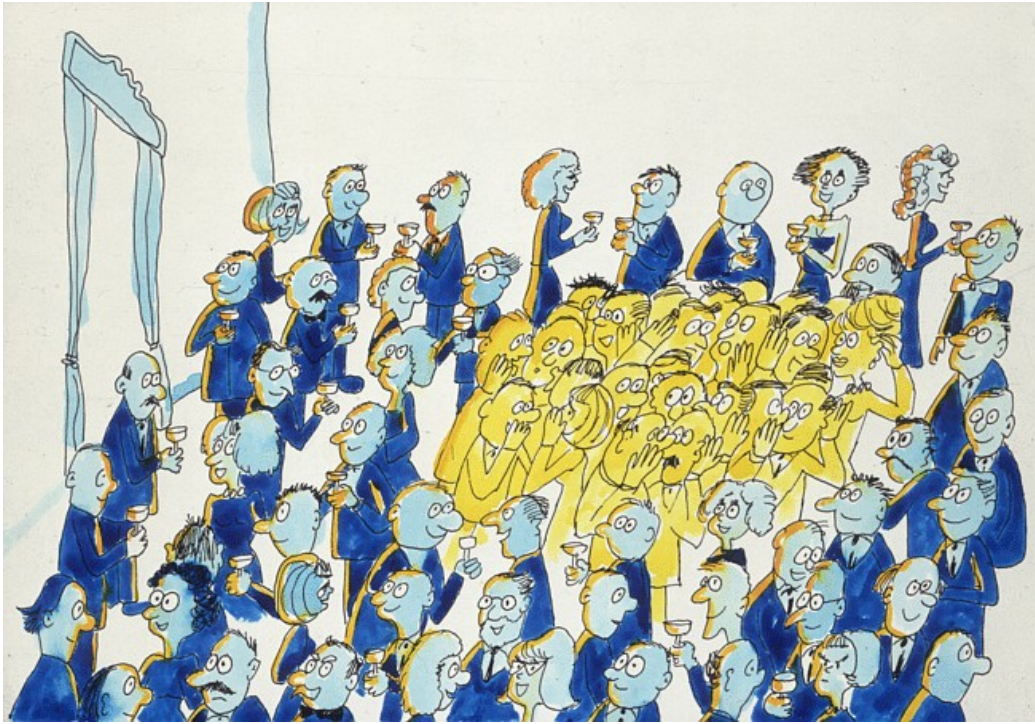

После того как мы намекнули, что нашей возможностью обсуждать эту статью мы обязаны смешиванию кварков, вас не должно удивить утверждение, что принцип суперпозиции не так чужд нашему сознанию, как обычно думают. Конечно, нам не приходилось видеть когерентную смесь живой и мертвой кошки. Но посмотрите на этот портрет Моны Лизы Джоконды

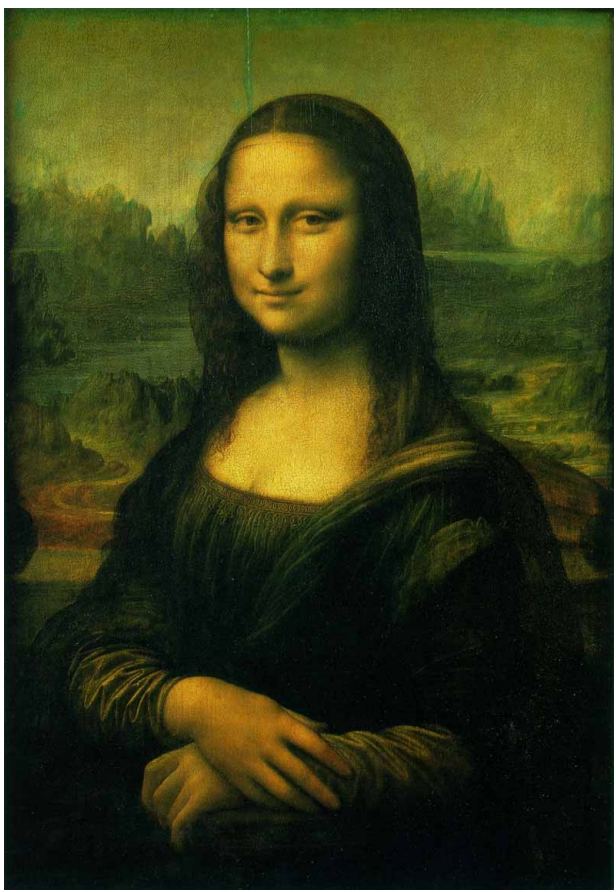

Обратите внимание на лицо женщины. Она как будто улыбается. Но в следующую минуту нам кажется, что она и вовсе не думала улыбаться, а тихо грустит. Выражение лица Моны Лизы как бы совмещает взаимно исключающие эмоции, как бы является их смесью. Вот вам и аналог квантовомеханической суперпозиции состояний [12]. Так вот, благодаря божественному таланту Леонардо да Винчи, вы узнали приятную новость: квантовую механику можно изучать в прелестном обществе женщин. Только учтите, не всякая бедная классическая мужская голова в состоянии долго терпеть квантовую логику женщин, их таинственную способность совмещать несовместимое.

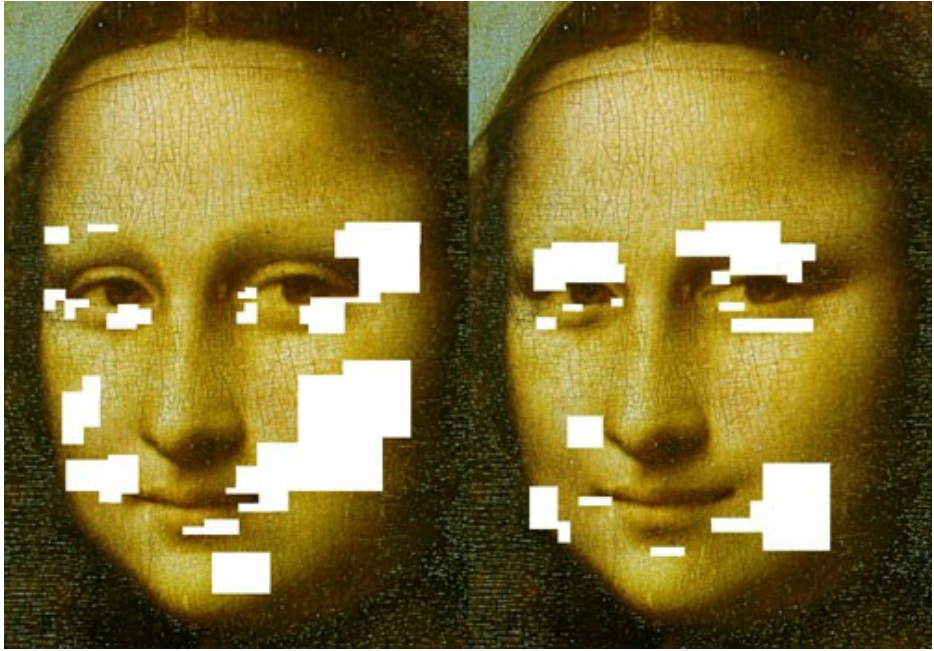

 В том, что выражение лица Моны Лизы действительно является смесью двух противоположных эмоции, можно убедиться если стереть определенные участки портрета [12], которые являются характерными для данной эмоции: На левом рисунке выражение лица явно грустное, тогда как на правом рисунке женщина беспечно улыбается.

Теперь, когда тайна улыбки Джоконды разгадана, нам осталось разгадать еще одну тайну. Мячик, брошенный вверх, движется по вполне определенной параболической траектории. Как же это согласуется с равноправием всех траекторий? Разгадка не так сложна, как может показаться. Представим себе, что фазы, ассоциированные с траекториями очень большие. Тогда величины $e^{is(P)} = \cos(s(P)) + i\sin(s(P))$ будут сильно осциллировать и вклады от разных траекторий взаимно сократятся. Но есть выделенный набор траекторий, которому это замечание не относится. Это траекторий, которые близки к экстремальной траектории $P_{cl}$, для которой $\delta s(P_{cl}) = 0$. То есть $P_{cl}$ выделена тем, что значение фазы не меняется (в первом порядке) при переходе к близкой к ней траектории. Тогда все траектории близкие к $P_{cl}$ будут интерферировать конструктивно, и они как раз и дадут главный вклад в амплитуду перехода. Вот и получили классическую траекторию $P_{cl}$. Ее квантовомеханическая "размазанность" тем меньше, чем больше соответствующие фазы. Заодно заметим, что мы получили принцип наименьшего действия $\delta s(P_{cl}) = 0$, который и определяет классическую траекторию. Только у нас $s(P)$ безразмерная величина, тогда как в классической механике действие $S(P)$ величина размерная. Следовательно, должны иметь $s(P) = S(P)/\hbar$, где $\hbar$ некая размерная константа, константа Планка, которая определяет масштаб квантового мира. Если характерные действия много больше, чем $\hbar$, то привычные представления классической механики можно с полным правом использовать. В противном случае будем иметь менее привычное квантовое поведение.

Принцип наименьшего действия является основой классической механики. Следовательно, классическая механика следует из квантовой механики, являясь ее предельным случаем, и в этом смысле менее фундаментальна.

Теперь об индетерминизме квантовой механики. Как следует из вышеизложенного, вероятностный характер лежит в самой основе квантовой механики. Это резко контрастирует с полным детерминизмом Ньютоновской механики. Надо признать, что это последнее нам более по душе. По-видимому, корни нашего предпочтения к детерминизму надо искать в первобытных пластах нашей души. В те далекие времена жизнь наших предков была полна опасности, но и, надо полагать, очарования, которое всегда сопровождает проявлениям спонтанности. Их мир был заселен духами, колдунами, шаманами и всякими предрассудками. С высоты сегодняшнего нашего относительного благополучия нам все это кажется глупостью. Но в те давнишние времена все это было жизненно важно для душевного благополучия наших предков. Было бы наивно думать, что весь этот тончайший психологический механизм приспособления, исправно служивший многие и многие тысячелетия, бесследно исчез из нашей психики, испугавшись тонкому слою просвещения последних веков. Если сверкают молнии, мы так же не сомневаемся, что это результат атмосферного электричества, как первобытный человек не сомневается в том, что это [13]:

" ...  ветреная Геба,

 Кормя Зевесова орла,

 Громокипящий кубок с неба,

 Смеясь, на землю пролила."

И еще неизвестно, какое объяснение по большому счету более правильное.

Детерминизм нужен первобытному человеку, чтобы чувствовать некоторую защищенность и духовный комфорт в этом изменчивом и враждебном мире. И в этом отношении мы принципиально не отличаемся от наших предков. Только  нам на помощь приходят законы Ньютона, а первобытному человеку – таинственные силы из мира духов. Послушаем Юнга [14]:

" Иногда эта гротескная идея принимает еще более впечатляющую форму. Так,  однажды один европеец застрелил крокодила, в желудке которого оказались два браслета. Туземцы признали в них браслеты, принадлежавшие двум женщинам, которых незадолго до этого проглотил крокодил. Тотчас же поднялся крик о колдовстве, так как этот вполне естественный случай, который не показался бы  подозрительным ни одному европейцу, был истолкован совершенно неожиданно, исходя из духовных предпосылок первобытного человека. Неведомый колдун вызвал крокодила, приказал поймать этих двух женщин и принести ему. Крокодил исполнил это приказание. Но как быть с двумя браслетами в желудке животного?

Крокодил, объясняли они, не проглатывает людей по собственной воле. Он получил браслеты в награду от колдуна.

Этот ценный случай является одной из иллюстраций произвольности объяснения в дологическом мире, видимо, потому, что нам такое объяснение представляется абсурдным и нелогичным. Но оно кажется нам таким лишь постольку, поскольку мы исходим из совершенно иных в сравнении с первобытным человеком предпосылок. Если бы мы подобно ему были убеждены в существовании колдунов и прочих таинственных сил, также как мы верим в так называемые естественные причины, то его вывод был бы для нас вполне логичным. На самом деле первобытный человек не более логичен или алогичен, чем мы. Просто он думает и живет исходя из совсем других предпосылок по сравнению с нами, В этом и состоит различие. Все, что выходит за рамки правил, все, что поэтому его беспокоит, пугает или удивляет, основывается для него на том, что мы называем сверхъестественным. Разумеется, он не считает это сверхъестественным, для него это является принадлежностью мира, который он познает. И если для нас естественно рассуждение типа: этот дом сгорел, потому что его поразила молния, то для первобытного человека столь же естественно сказать: колдун воспользовался молнией, чтобы поджечь именно этот дом. В мире первобытного человека абсолютно все, что сколько-нибудь необычно или впечатляюще, подвергается подобному либо принципиально сходному объяснению. Но при этом он поступает точно так же, как и мы: он не задумывается над своими исходными посылками. Для него несомненно a priori, что болезнь и т.д. вызывается духами или колдовством, также как для нас с самого начала не вызывает сомнений, что болезнь имеет так называемую естественную причину. Мы столь же мало думаем о колдовстве, как он о естественных причинах. Само по себе его духовное функционирование принципиально не отличается от нашего. Различаются, как я уже говорил, только исходные посылки."

Но надо сказать, что если целью является обеспечение предсказуемости окружающего нас мира и связанный с этим некоторый духовный комфорт, мы со своими законами Ньютона просчитались больше, чем первобытный человек со своими колдунами. Увы, детерминизм классической физики не сможет обеспечить нам предсказуемость будущего больше, чем оракулы прошлого. Конечно, я утрирую, и это не совсем так. С простыми и искусственными системами и ситуациями мы вполне можем справиться. Нас окружают механизмы, которые вполне предсказуемо функционируют. Но, например, долгосрочный прогноз погоды мы таким же успехом можем спросить у Дельфийского оракула, как у метеорологов. И для этого есть веские причины.

Дело в том, что классическая механика точно предсказывает будущее только, если точно известны начальные условия: начальные координаты и скорости всех частиц. Но мы не можем абсолютно точно знать начальные условия. До примерно 1960 г., общепринятым было мнение, что это не имеет принципиального значения: если немного ошибаемся в начальных условиях, наше предсказание будет неточным, но разница будет невелика и значит не существенная. Если требуется больше точности в предсказаниях, мы всегда можем более точно измерить начальные условия. Все изменилось, когда американский метеоролог Эдвард Лоренц при компьютерном моделировании погоды случайно открыл замечательное явление, которое впоследствии получило название "эффект бабочки" [15]. Образно говоря, один взмах крыльев бабочки сегодня в Бразилии через месяц может вызвать торнадо в Калифорнии. Более научно, это называется развитием хаоса, когда незначительные изменения начальных условии приводят к радикально разным последствиям. Интересно, что существование подобных непредсказуемых динамических систем предполагал еще Пуанкаре [16]:

"Если бы мы знали точно законы природы и состояние Вселенной в начальный момент, то мы могли бы точно предсказать состояние Вселенной в любой последующий момент. Но даже и в том случае, если бы законы природы не представляли собой никакой тайны, мы могли бы знать первоначальное состояние только приближенно. Если это нам позволяет предвидеть дальнейшее ее состояние с тем же приближением, то это все, что нам нужно. Мы говорим, что явление было предвидено, что оно управляется законами. Но дело не всегда обстоит так; иногда небольшая разница в первоначальном состоянии вызывает большое различие в окончательном явлении. Небольшая погрешность в первом вызвала бы огромную ошибку в последнем. Предсказание становится невозможным, мы имеем перед собой явление случайное."

Но эти идеи Пуанкаре не получили тогда должного развития. Наверно считалось, что подобные непредсказуемые системы нечто исключительные, и в основном детерминизм все равно торжествует. После открытия Лоренца, постепенно стало очевидным, что верно как раз обратное. Хаос, чрезмерная чувствительность к начальным условиям, является правилом, а не исключением для реальных систем. Даже Солнечная система хаотична, и мы не можем точно предсказать положение планет через сотни миллионов лет. Например, импульс от запуска ракеты может привести к повороту Земли на своей орбите [17] на 60 градусов через двести миллионов лет.

Но это еще не все. Пусть у нас есть сосуд с газом и представим невозможное: мы точно знаем начальные координаты и скорости всех молекул в этом сосуде. Тогда классическая механика позволяет предсказать дальнейший путь одной какой-нибудь молекулы точно. Но это справедливо, только если мы сможем изолировать наш сосуд с газом от внешнего воздействия. Есть одно воздействие, гравитация, которое невозможно экранировать. Так вот, представим, что во Вселенной ничего нет кроме нашего сосуда с газом и одного единственного электрона на краю видимой части Вселенной, то есть на расстоянии десять миллиардов световых лет от сосуда с газом. Трудно себе представить влияние менее незначительное, чем влияние этого электрона на молекулы газа. Тем не менее, это делает сосуд с газом не замкнутой системой, и мы вправе ожидать, что направление движения выделенной нами молекулы, в конце концов, отклонится существенно от предсказания. Весь вопрос в том, через сколько времени это произойдет. Если ваша интуиция подсказывает вам, что на это уйдут миллиарды и миллиарды лет, то она жестоко вас обманывает. Это произойдет очень быстро, меньше чем за одну миллионную долю секунды мы полностью потеряем предсказательную способность из-за гравитационного притяжения одного единственного электрона на краю Вселенной! [18].

Это похоже на чудо, но на самом деле все очень просто. Механизм, который способен за короткое время чрезмерно усилить маленькое возмущение, был известен еще Пуанкаре [16]: "Если бы молекула уклонилась налево или направо от своей траектории на очень малую величину, сравнимую с радиусом действия молекул газа, то она избежала бы толчка или таковой произошел бы при совершенно иных условиях, а это могло бы изменить на 90 или 180 градусов направление скорости после толчка. И это еще не все. Как мы видели, достаточно отклонить молекулу до толчка на бесконечно малое расстояние, чтобы она после толчка отклонилась на конечное расстояние. Поэтому, если бы молекула подверглась двум последовательным столкновениям, то ей достаточно было бы сообщить до первого толчка бесконечно малое уклонение второго порядка, чтобы мы получили после первого столкновения бесконечно малое уклонение первого порядка, а после второго - конечное. Между тем молекула испытывает не только два столкновения, а весьма большое число их в секунду. Поэтому, если первый толчок умножает отклонение на весьма большое число $A$, то после $n$ столкновений оно будет умножено на $A^n$. Оно сделается, следовательно, весьма большим не только потому, что $A$ очень велико, т.е. потому, что малые причины производят большие следствия, но и потому, что показатель n велик, т.е. потому, что столкновения весьма многочисленны и причины очень сложны."

Гравитационное притяжение электрона приводит к тому, что молекула немного отклоняется от пути перед первым столкновением. Ясно, что отклонение совершенно мизерное, что то вроде $10^{-104}$. Но каждое столкновение увеличивает неопределенность в направлении движения молекулы в сотни раз. Поэтому достаточно всего лишь около 50 столкновений, чтобы исходная мизерная неопределенность превратилась в величину порядка единицы.

Любопытно отметить, что в свете этого ошеломляющего примера влияния электрона с края Вселенной, уже не кажется таким уж фантастическим предположение Бергсона [19], что "функция мозга, нервной системы и органов чувств, в основном, выделительная, а не продуктивная. Каждая личность в каждый момент способна помнить все, что когда-либо с нею происходило, и воспринимать все, что происходит везде во вселенной. Функция мозга и нервной системы заключается в том, чтобы защитить нас от этой массы, в основном, бесполезного и не имеющего смысла знания, ошеломляющего и повергающего нас в смятение, исключая большую часть того, что иначе мы бы воспринимали и помнили в любой момент, и оставляя лишь очень маленькую и особую подборку того, что, вероятнее всего, окажется практически

полезным."

Таким образом, квантовая механика, хотя и по-новому ставит интересный и глубокий философский вопрос о существовании и смысле случайных событий, вряд ли метафизически менее подходящая для нас, чем классическая механика. Можно сказать, что даже более подходящая, так как ближе к архаической установке первобытного человека (особенно если принять, что сознание и вправду играет роль в редукции волновой функции).

Но вероятность в квантовой механике входит как-то по-особому: складываются амплитуды вероятностей, а не сами вероятности. Я не буду прикидываться, что второе и третье основные правила движения электрона также "очевидны", как первое правило. Они, безусловно, немного сложноваты, им не хватает самоочевидности аксиом. По аналогии с аксиомой параллельности в геометрии, тогда встает вопрос об их единственности. Например, можно ли вместо комплексных чисел использовать какие-либо другие числа?

Представим себе, что пространство между двумя точками разделено на две части непроницаемым экраном с дыркой посередине. Пусть амплитуда перехода из первой точки к отверстию экрана $z_1$, а из отверстия к второй точке - $z_2$. Так как действие аддитивная величина вдоль траектории, то амплитуда перехода из первой точки во вторую равна $z_1 z_2$, а вероятность $|z_1 z_2|^2$. Но вероятность такого перехода равна вероятности последовательного наступления двух событий: электрон оказывается около отверстия, вероятность этого равна $|z_1|^2$, и электрон из отверстия оказывается во второй точке с вероятностью $|z_2|^2$. Следовательно, мы должны иметь $|z_1 z_2|^2 = |z_1|^2 |z_2|^2$. Это означает, что числовая система, которая может заменить комплексные числа, должна составлять композиционную алгебру. И у нас не так много альтернатив. Даже совсем мало: комплексные числа можем заменить только на кватернионы или октонионы [20]. Кватернионную квантовую механику развивали довольно подробно [21], но пока это не привело к скольким-нибудь значительным открытиям, хотя бы отдаленно сравнимыми с появлением (комплексной) квантовой механики или неэвклидовой геометрии.

Любопытно, что квантовая механика даже ограничивает развитие хаоса. Представим себе некоторую хаотичную систему. Так как мы не можем точно знать ее начальное положение, то этому начальному положению будет соответствовать некая область конечного объема в фазовом пространстве, а не одна точка. Из-за предполагаемой хаотичности системы, точки из этой начальной области будут со временем экспоненциально расходиться. При этом объем области будет сохраняться (так называемая теорема Лиувилля). Образно, область начальных состояний можно представить как каплю жидкости. Со временем эта капля будет все больше и больше дробиться, дочерние капельки будут все мельче и мельче, и они будут рассеяны по всему фазовому пространству. Хотя при этом суммарный объем капелек не меняется, их суммарная поверхность будет расти экспоненциально. Характерное время этого экспоненциального роста есть характерное время хаоса: время в течение которого система полностью забывает свое начальное состояние. Это - наглядная картина классического хаоса. Что нового привносит квантовая механика в эту картину? Из-за соотношения неопределенности Гейзенберга в фазовом пространстве квантовой системы есть предел дробления. Следовательно, через некоторое время, когда все дочерние капельки приобретут минимальный объем, допустимый соотношением неопределенности, развитие хаоса прекратится. Интересно, насколько быстро это произойдет? Рассмотрим конкретный пример - хаотичное вращение Гипериона [22], шестнадцатого спутника Сатурна.

Несферическая форма Гипериона, гравитационное притяжение самого Сатурна и резонансное влияние его большого спутника Титана приводят к тому, что Гиперион хаотически кувыркается. При этом характерное время развития хаоса, примерно 100 дней, много больше, чем его орбитальный период (21.3 дня) и его перед вращения (5 дней). Трудно сомневаться, что Гиперион классическое тело. Момент импульса этой огромной "картофелины" (с размером около 142 км) примерно в $10^{58}$ раз больше, чем элементарный Планковский квант момента импульса. Поэтому можно предположить, что квантовое насыщение хаоса в этом случае наступает через космологически большое время и, следовательно, не имеет никакого значения. Правда, однако, состоит в другом [22]: отношение этого самого времени квантового насыщения хаоса к времени развития хаоса равно логарифму от приведенного выше большого числа $10^{58}$, что уже не такое уж большое число. В результате получается, что квантовое насыщение хаоса во вращении Гипериона должен наступить всего лишь через 37 лет! Но тогда возникает вопрос: существует ли классический хаос вообще? Ответ утвердительный. Дело в том, что квантовый эффект подавления хаоса перечеркивается другим квантовым эффектом - де-когерентностью из-за присутствия внешнего случайного возмущения. Например, солнечный свет, который падает на Гиперион и вряд ли может повлиять на его основную динамику, тем не менее, вносит достаточное возмущение, чтобы привести к потере квантовой когерентности и к восстановлению классического хаоса [22]. Опять очень маленькие возмущения (помните электрон на краю Вселенной?) оказываются важными, чтобы получить правильную картину явления.

На тот случай, если все мои усилия были тщетны, и вы все равно сомневаетесь в том, что квантовая механика в своей основе очень проста, естественна и глубока, я приберег еще два аргумента в пользу того, что электрон действительно использует все возможные пути. Первый аргумент - это эффект Ааронова-Бома.

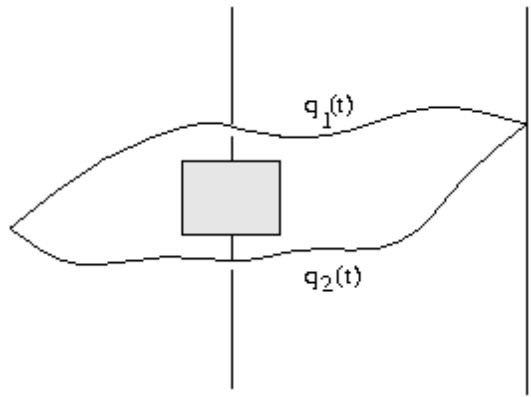

Пусть на пути электрона имеется непроницаемый экран с двумя щелями, а между ними непроницаемая область, где можно создать магнитное поле. На классический электрон это поле никак не подействует, так как у электрона нет возможности оказаться в области поля. Этой возможности нет и у квантового электрона, но из-за того, что он использует все доступные пути, возникает любопытный эффект. Во-первых, даже при отсутствии магнитного поля мы будем наблюдать интерференционную картину на втором экране. Это легко объяснить, так как вероятность попадания электрона в данную точку второго экрана равна $|z_1 + z_2|^2$, где первое слагаемое соответствует путям, которые проходят через первое отверстие, а второе слагаемое путям, которые проходят через второе отверстие. В зависимости от относительной фазы комплексных чисел $z_1$ и $z_2$, эта величина будет меняться.

Но самое интересное, что при включении магнитного поля интерференционная картина сдвигается! Дело в том, что действие и, следовательно, фаза, соответствующая данному пути, зависит не от магнитного поля, а от вектор-потенциала, который отличен от нуля и вне запрещенной зоны. Это можно увидеть так. Рассмотрим циркуляцию вектор-потенциала по замкнутому контуру, составленному из двух указанных на рисунке путей. По теореме Стокса [23], этот интеграл равен потоку ротора вектор-потенциала (т.е. магнитного поля) через поверхность ограниченной замкнутым контуром. Следовательно, разность фаз между двумя путями, проходящими через разные щели, определяется только магнитным потоком через запрещенную область и вовсе не зависит от конкретного вида путей [24]. Поэтому при включении магнитного поля амплитуды $z_1$ и $z_2$ приобретут дополнительную, зависящую только от магнитного потока, различность фаз и интерференционная картина сдвинется.

Этот пример показывает, что понятие силы не является фундаментальным в квантовой механике: есть ситуации, когда на частицу не действуют никакие силы, а эффект все равно присутствует. Это непривычно для нашей интуиции, воспитанной на классической физике, но эксперименты показывают, что эффект действительно наблюдается [25].

Второй пример должен быть весьма убедительным: то, что мы не проваливаемся через пол. Вас это не удивляет? Ведь атом практически пустой - атомное ядро составляет крохотную часть размера атома. Почему тогда атомы, из которых мы состоим, не проходят довольно свободно через атомы составляющих пол? Чтобы понять это, надо вспомнить, что все электроны одинаковы — невозможно отличить один какой-нибудь электрон от другого. Поэтому, при распространении двух электронов возможны два класса путей:

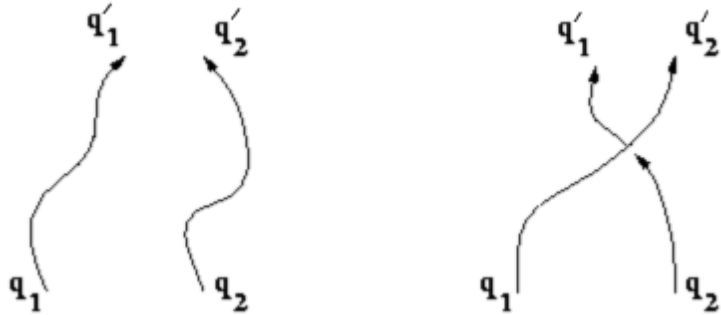

Оба класса, однако, приводят к одной и той же конечной ситуации, так как пути из второго класса соответствуют перестановке идентичных электронов. Если бы электроны не были идентичны, каждый класс путей соответствовал бы разным амплитудам перехода, и мы их не складывали бы. Но сейчас они приводят к одному и тому же результату, т.е. мы их вроде должны сложить. Но при этом мы не может быть уверены, что при перестановке электронов не возникнет дополнительный фазовый множитель: ведь с одной стороны вроде пути из двух классов равноправны, а с другой стороны они все-таки отличаются. Поэтому общий вид амплитуды перехода для двух электронов будет [24] $K(q_1,q_2;q_1',q_2') = A(q_1,q_2;q_1',q_2') + e^{i\phi} A(q_1,q_2;q_2',q_1')$ . Но амплитуда $K(q_1,q_2;q_2',q_1')$ соответствует тому же самому конечному состоянию. Поэтому он может отличаться от $K(q_1,q_2;q_1',q_2')$ только по фазе: $K(q_1,q_2;q_2',q_1') = e^{i\alpha} K(q_1,q_2;q_1',q_2')$ . Вспоминая выражение для амплитуды $K$, это дает $\alpha = \phi$ и $e^{2i\phi} = 1$. Следовательно, или $\phi = 0$, или $\phi = \pi$. В первом случае говорят, что идентичные частицы удовлетворяют статистике Бозе-Эйнштейна и амплитуды из двух классов путей складываются. Во втором случае имеем так называемую статистику Ферми - Дирака, амплитуды из двух классов путей вычитаются, и два электрона не могут быть в одном и том же состоянии (принцип запрета Паули - амплитуда перехода для такой конфигурации зануляется). То, что мы не проваливаемся через пол, указывает, что электроны подчиняются статистике Ферми - Дирака: электронные облака атомов перекрываются неохотно, т.е. как бы возникает сила отталкивания при сближении атомов. Таким образом, твердость тел - тоже весьма тонкий квантовый эффект.

# Примечание A:   Комплексные числа

Данную историю [26] рассказала своему коллеге Абрааму Гольдбергу методист кабинета математики Областного института усовершенствования учителей. Она инспектировала сельскую школу одного из районов Ленинградской области. В школе преподавателей не хватало и зимой по некоторым предметам их замещал молодой тракторист. Инспектор пришла на урок математики. Тракторист вел урок уверенно и блистал сообразительностью. Детям он нравился, и они охотно отвечали на его вопросы, демонстрирую свои знания. Дошло дело до нового материала: сложение правильных дробей. Преподаватель по совместительству объяснил, что надо складывать числители отдельно, потом знаменатели отдельно и получить результат. Инспектор не поверила своим ушам, но не стала при всем классе указывать молодому человеку на его ошибки и только после урока объяснила ему на нескольких примерах почему правильные дроби надо складывать иначе. Парень пообещал на следующем уроке все исправить.

На следующем уроке, после проверки домашнего задания, тракторист похвалил детей и сказал, что так складывали дроби до революции 1917 года, а после революции их складывают иначе и объяснил новые правила. Дети труднее, но все таки поняли и этот способ, только один второгодник, сидящий рядом с инспектором на задней парте, громко посетовал как сложно стали складывать дроби после революции.

Однажды несколько искаженную эту историю в качестве анекдота рассказали в обществе знаменитого математика Гельфанда [27]. Один из математиков, «присутствовавший при разговоре, откликнулся на анекдот неожиданным для собеседников пассажем:

– Ну, а почему, собственно, нельзя так складывать дроби? Давайте введем такое правило сложения дробей и так и будем их складывать! Что нам мешает?

На это сейчас же последовало возражение:

– Но ведь тогда не будет выполняться распределительный закон!

– Ну если при этом всё остальное оставить без изменения, то конечно! А мы изменим правило умножения дробей так, чтобы распределительный закон выполнялся, как и все остальные свойства сложения и умножения.

– Ну, ну, попробуйте! – еще больше удивились собеседники. – Интересно, как это Вам удастся!

В ответ на доске появилась формула: $\frac{a}{b} \cdot \frac{c}{d} = \frac{ac-bd}{bc+ad}$.

Вот тут Гельфанд и засмеялся:

– Здорово! – сказал он. – Напишите про это.»

Гельфанд засмеялся, потому что узнал в формуле правило умножения комплексных чисел. Через 15 лет собеседник Гельфанда выполнил его поручение, и так появилась занимательная статья [27] про комплексные числа, которую советуем посмотреть тем, которые не знают, или не помнят, что такое комплексное число.

Здесь только отметим, что правило сложения "дробей" $\frac{a}{b}+\frac{c}{d}=\frac{a+c}{b+d}$ напоминает сложение векторов (a,b) и (c,d). Мы можем представить себе комплексное число как конечную точку этого вектора (а начальная точка всегда находится в начале координат). Тогда каждому комплексному числу (x,y) можно соотнести его модуль $r=|(x,y)|=\sqrt{x^2+y^2}$ – длину этого вектора, и фазу $\varphi$ – угол, который вектор (x,y) составляет с осью x, см. рисунок:

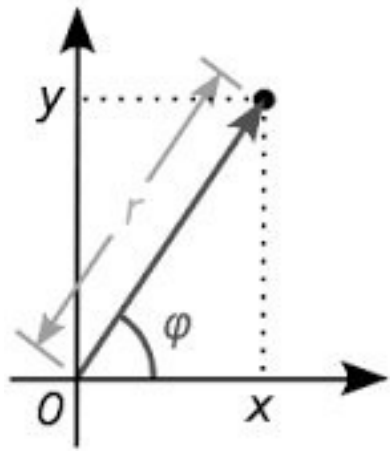

## Примечание B: Смешивание кварков

Графически распространение массивной частицы можно представить как последовательность элементарных актов взаимодействия безмассовой частицы с вакуумом:

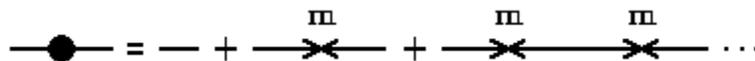

Интенсивность элементарного акта взаимодействия пропорциональна массе частицы.

Здесь мы должны предупредить вас, чтобы вы не воспринимали все это слишком буквально. Точный смысл всем этим словам можно придать только в рамках квантовой теории поля (см., например, М. Пескин, Д. Шредер, Введение в квантовую теорию поля. Москва, РХД, 2001). В частности, вакуум не всегда означает пустоту, так как под вакуумом понимается состояние с наименьшей энергией, а это не всегда соответствует отсутствию полей.

Представим себе, что безмассовая частица (кварк) может взаимодействовать со скалярным полем: излучать скалярный квант подобно тому, как электрон может излучать квант электромагнитного поля - фотон. Схематически это можно изобразить так ($g$ характеризует интенсивность взаимодействия - аналог заряда электрона):

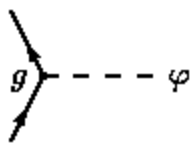

Но самодействие скалярного поля может быть устроено так, что состоянием с наименьшей энергией будет не пустота, т.е. отсутствие поля, а состояние в котором скалярное поле имеет некое отличное от нуля значение (вакуумное среднее) $\langle \phi \rangle$. В этом случае, скалярное поле удобно разложить так $\phi = \langle \phi \rangle + \phi'$. Тогда вершина нашего элементарного взаимодействия разобьётся на две части:

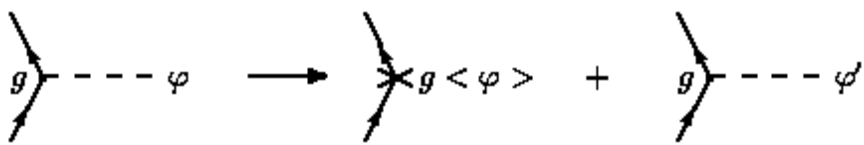

Вторая из них соответствует излучению кварком реальной скалярной частицы (Хиггсовского бозона), а первая генерирует массу кварка $m = g \langle \phi \rangle$.

Но если кварк может испустить скалярную частицу без изменения своего типа (аромата), почему он не может сделать это с изменением аромата? Ведь мы знаем, что тип кварков не всегда сохраняется. Кварковый аромат меняется, например, в слабых взаимодействиях. Представим себе, что d-кварк при излучении скалярного кванта переходит в s-кварк. К каким последствиям приведет тогда существование скалярного конденсата в вакууме? Выделим из соответствующей вершины взаимодействия вакуумное среднее скалярного поля:

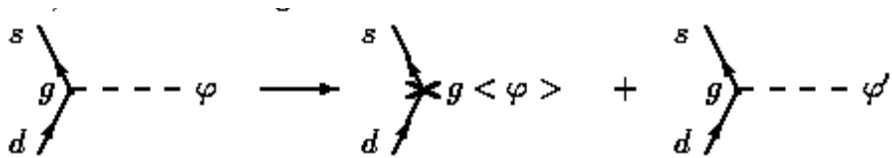

Но теперь первый член приводит к d-s смешиванию! В результате d и s-кварки перестанут иметь определенные массы, и только их линейные комбинации (физические d и s-кварки) будут иметь определенные массы. В данном случае смешивание характеризуется одним параметром - углом Каббибо:

$$\begin{pmatrix} \tilde{d} \\ \tilde{s} \end{pmatrix} = \begin{pmatrix} \cos\theta_c & -\sin\theta_c \\ \sin\theta_c & \cos\theta_c \end{pmatrix} \begin{pmatrix} d \\ s \end{pmatrix} \longrightarrow \begin{array}{l} \tilde{d} = \cos\theta_c\, d - \sin\theta_c\, s \\ \tilde{s} = \sin\theta_c\, d + \cos\theta_c\, s \end{array}$$

Это смешивание приводит к наблюдаемым последствиям. Если изначально слабые взаимодействия были возможны только в рамках одного поколения кварков, т.е. внутри пар (u,d) и (c,s) "голых" кварков, то после смешивания слабые переходы между разными поколениями тоже станут возможными, так как физический s-кварк, например, содержит в себе обе "голые" s и d кварки. Как следствие, слабый переход физического u-кварка в физический s-кварк будет пропорционален синусу угла Каббибо - малой величине.

Но в Природе существует три поколения кварков: (u,d), (c,s), (t,b). После смешивания, любой "верхний" кварк u,c,t может переходить в любой "нижний" кварк d,s,b в результате слабого взаимодействия. Следовательно, интенсивность слабых переходов описывается с помощью 3X3 матрицы Кобаяши-Маскава, которая является обобщением угла Каббибо. Экспериментально установлено, что матрица Кобаяши-Маскава имеет иерархическую структуру, и ее приблизительно можно параметризовать так:

$$\begin{pmatrix} V_{ud} & V_{us} & V_{ub} \\ V_{cd} & V_{cs} & V_{cb} \\ V_{td} & V_{ts} & V_{tb} \end{pmatrix} \approx \begin{pmatrix} 1 - \frac{1}{2}\lambda^2 & \lambda & A\lambda^3(\rho - i\eta) \\ -\lambda & 1 - \frac{1}{2}\lambda^2 & A\lambda^2 \\ A\lambda^3(1 - \rho - i\eta) & -A\lambda^2 & 1 \end{pmatrix}$$

Здесь $\lambda$ является малым параметром. Следовательно, слабые переходы самые интенсивные внутри одного и того же поколения. Переходы между поколениями, хоть и присуствуют, но они подавлены. Причем, переходы между удаленными поколениями подавлены более сильно, чем переходы между соседними поколениями.

Если $\eta \neq 0$, то матрица Кобаяши-Маскава комплексная, и CP симметрия нарушается. Это очень удивительно, так как нарушение CP означает, что или Природа различает правое и левое, или существует целый зеркальный мир [28]. Но эксперименты показывают, что дело обстоит именно так: CP нарушается в распадах К-мезонов, В-мезонов. И слава Богу, что CP нарушается, так как иначе нас бы не было! Опираясь на идею Сахарова [29] (см. также здесь [30], сейчас общепринято, что возникновение барионной асимметрии Вселенной требует CP-нарушения, как одного из условий. Иначе частицы и античастицы образовались бы в равных количествах после Большого Взрыва, они бы аннигилировались полностью, и во Вселенной не было бы вещества.

# Список цитированной литературы